\documentclass[aps,prl,reprint]{revtex4-2}

\usepackage{graphicx}
\usepackage{dcolumn}
\usepackage{bm}
\usepackage{amsmath,amssymb}
\usepackage{mathrsfs}
\usepackage[caption=false]{subfig} 
\usepackage{braket}

\usepackage{xcolor}

\definecolor{linkcol}{RGB}{18, 94, 173}     
\definecolor{citecol}{RGB}{166, 45, 124}    
\definecolor{urlcol}{RGB}{0, 128, 112}      

\usepackage[
    colorlinks=true,
    linkcolor=linkcol,
    citecolor=citecol,
    urlcolor=urlcol,
]{hyperref}

\newcommand\hphi{\hat{\phi}}
\newcommand\mreg{\mathcal{M}_\text{reg}}
\newcommand\mrcmps{\mathcal{M}}
\newcommand\tr{\mathrm{tr}}
\newcommand\upd{\text{d}}
\newcommand\e{\mathrm{e}}
\DeclareMathOperator*{\argmin}{argmin} 

\newcounter{appsec}
\renewcommand{\theappsec}{\Alph{appsec}}   

\begin{document}

\title{Multi-Field Relativistic Continuous Matrix Product States}

\author{Karan Tiwana}
\email{karantiwana21693@gmail.com}
\author{Antoine Tilloy}
\email{antoine.tilloy@minesparis.psl.eu}
\affiliation{Laboratoire de Physique de l'École Normale Supérieure, Mines Paris - PSL, Inria, CNRS, ENS-PSL, Sorbonne Université, PSL Research University, Paris, France}

\date{\today}

\begin{abstract}
\noindent Relativistic continuous matrix product states (RCMPS) are a powerful variational ansatz for quantum field theories of a single field. However, they inherit a property of their non-relativistic counterpart that makes them divergent for models with multiple fields, unless a regularity condition is satisfied. This has so far restricted the use of RCMPS to toy models with a single self-interacting field. We address this long standing problem by introducing a Riemannian optimization framework, that allows to minimize the energy density over the regular submanifold of multi-field RCMPS, and thus to retain purely variational results. We demonstrate its power on a model of two interacting scalar fields in $1+1$ dimensions. The method captures distinct symmetry-breaking phases, and the signature of a Berezinskii-Kosterlitz-Thouless (BKT) transition along an $O(2)$-symmetric parameter line. This makes RCMPS usable for a far larger class of problems than before.
\end{abstract}

\maketitle
\emph{Introduction} --- Tensor network states are a powerful ansatz for the non-perturbative study of quantum many-body systems~\cite{SCHOLLWOCK201196, ORUS2014117}, and have allowed tremendous progress in the understanding of lattice problems. Strongly coupled quantum field theories (QFT) provide particularly extreme instances of the many-body problem, and are still challenging to solve, even in low dimensions. One natural approach is to first discretize, and study the resulting lattice model with $1$ dimensional tensor networks (matrix product states, or MPS). In this context, MPS provide precise results in the non-perturbative regime \cite{milsted2013_MPSforPhi4,Banuls:2016gid}, which need to be carefully extrapolated numerically to the continuum.

A possibly more appealing approach is to take the continuum limit analytically first, with the Continuous Matrix Product State (CMPS) \cite{PhysRevLett.104.190405,Haegeman2013}. The latter are natively adapted to the study of non-relativistic QFT. However, when used as a variational ansatz for a relativistic QFT Hamiltonian, CMPS still require an extra UV cutoff~\cite{PhysRevLett.105.251601,stojevic2015entanglementscaling}, and thus do not fully remove the need for an extrapolation. More recently, CMPS have been combined with a Gaussian disentanglement (or Bogoliubov) transform \cite{vardian2023rcmps_as_cmera}, to yield a new ansatz, the relativistic CMPS (RCMPS) ansatz~\cite{Tilloy:2021hhb, Tilloy:2021yre}. RCMPS can be used as a variational ansatz directly in relativistic QFT, without the need to introduce any cutoff, UV or IR. As such, they embody Feynman's hope for the variational method in QFT~\cite{Feynman1987}, at least in low dimension.

In $1+1$ dimensions, RCMPS can deal with most single-field relativistic bosonic models. Such models are defined by their potential $V(\hat{\phi})$, which can take the form of any (classically) lower-bounded polynomial or exponential. In this low-dimensional context, renormalization is particularly simple: normal-ordering the potential $V(\hat{\phi}) \rightarrow :\!V(\hat{\phi})\!:$ is sufficient to remove \emph{all} UV divergences and to make the model well defined \footnote{Normal-ordering is sufficient for all monomials $\hat{\phi}^n$, and all exponentials $\exp(i\alpha\hat{\phi})$ with $\alpha \leq \sqrt{4\pi}$, with the definition of the scalar field that we give in the text.}. The Hamiltonian density is then lower-bounded, which provides a well defined variational problem~\cite{glimm1968_phi4bound, federbush1969_phi4bound}. As a result, RCMPS have already been used to study the low energy properties of the $\phi^4$ \cite{Tilloy:2021yre}, Sine-Gordon, and Sinh-Gordon models \cite{Tilloy:2022kcn}, and have allowed to compute all local and some extended operators~\cite{Tiwana:2025ptw} in the ground state.

However, many interesting phenomena in QFT, from flavor symmetries to topological phases, arise in theories with multiple fields \cite{Peskin:1995ev, Fradkin_2013, Zee:2003mt}. Even the simplest fermionic model, the free Dirac field, is a multi-component field composed of particle and anti-particle field operators. Extending the CMPS and RCMPS toolbox to multi-field systems introduces a major technical challenge, which has so far remained unaddressed. It appears already in the case of two scalar fields. For such systems, the RCMPS ansatz is parametrized by $3$ matrices $Q$, $R_1$, and $R_2$. For generic values of the $R_1,R_2$ matrices, the kinetic part of the energy density is infinite, a problem that does not occur in the single field case. To obtain a finite energy density, and thus a well defined variational problem, the matrices $R_1,R_2$ must satisfy a \emph{regularity conditions} $[R_1, R_2] = 0$ (see \cite{Haegeman2013} for proof in the case of CMPS, and End Matter \ref{app:regularity} for RCMPS). This restricts the energy minimization to a non-trivial submanifold $\mreg$ within the full RCMPS space $\mrcmps$, and makes standard gradient minimization impossible.

In this Letter, we introduce a Riemannian optimization algorithm that allows us to iteratively minimize the energy density of any model on the space $\mreg$ of regular multi-field RCMPS. Starting from a point $p_k \in \mreg$, each iteration consists of two steps: (1) an orthogonal projection of the regular part of the energy density gradient onto the tangent space $T_p\mreg$ of $\mreg$, which provides the steepest descent direction, and (2) a retraction along this descent direction, that provides a point $p_{k+1}$ on $\mreg$ even for large steps. Each iteration gets the state closer to the ground state, without introducing divergent contributions, and we may stop when the projected gradient is small enough. We illustrate the power of this method on the example of a coupled scalar field model with dihedral group ($D_4$) symmetry. RCMPS allow us to get results far more precise than third order perturbation theory, probe deep into non-perturbative parts of the phase diagram, and even witness the signs of a Berezinskii-Kosterlitz-Thouless (BKT) transition.

\emph{An Example Model and the Regularity Constraint.}---To make the discussion explicit, we consider a model of two coupled and self-interacting scalar fields  $\hphi_1$ and $\hphi_2$ with Hamiltonian $H = H_1^{0} + H_2^{0} + V(\hphi_1,\hphi_2)$. The free Hamiltonians are defined as usual as
\begin{equation}\label{eq:freepart}
  H_j^{0} = \int_\mathbb{R} \upd x  :\frac{\hat{\pi}_j^2}{2} + \frac{(\partial_x\hat{\phi}_j)^2}{2} + \frac{m^2}{2}\hat{\phi}_j^2: 
\end{equation}
where the operators $\hat{\pi}_j$ are the momenta canonically conjugate to $\hphi_j$.
For the potential, we choose:
\begin{equation}\label{eq:potential}
  V(\hphi_1,\hphi_2) = \int_\mathbb{R}\upd x  g \,\big(:\!\hat{\phi}_1^4\!:+:\!\hat{\phi}_2^4\!:)+ \,2\lambda:\hat{\phi}_1^2\hat{\phi}_2^2: \, ,
\end{equation}
which is a quite general quartic potential. It is possible to choose different $m$ and $g$ for the different fields, but the present restriction gives non-trivial symmetries, leading to interesting physics. 

In these definitions, the normal-ordering is defined with respect to the normal modes $\hat{a}_{j,p}, \hat{a}_{j,p}^\dagger$ that diagonalize the free part \eqref{eq:freepart}. The latter are related to the field operators $\hat{\phi}_j$ with the standard mode expansion: 
\begin{equation} 
\hphi_j(x) = \frac{1}{2\pi}\int_\mathbb{R} \frac{\upd p}{\sqrt{2\omega_{p}}} (\hat{a}_{j,p}e^{ipx} + \hat{a}^{\dagger}_{j,p}e^{-ipx})~, \end{equation} 
with $\omega_p = \sqrt{p^2+m^2}$. With this choice of normal ordering (equivalent to tadpole cancellation), the Hamiltonian $H$ is well defined non-perturbatively, and has a lower bounded ground state energy density for $g>0$.

An important object in the RCMPS construction is the Fourier transform of the normal modes~\cite{Tilloy:2021hhb}:
\begin{equation}
  a_j(x) := \frac{1}{2\pi} \int_{\mathbb{R}} \upd p\, a_{j,p} \e^{ipx} \, .
\end{equation}
These spatial modes are related to the field $\hphi_j(x)$ non-locally, \textit{i.e.} via  a convolution:
\begin{equation}
  \hphi_j(x) = \int_\mathbb{R} \upd y\, J(x-y) [\hat{a}_j(x) + \hat{a}^{\dagger}_j(x)],
\end{equation}
where
$J(x) = \frac{1}{2\pi} \int_\mathbb{R} \frac{\upd k}{\sqrt{2\omega_k}}e^{-i k x}$ is a function that decays exponentially as $x\rightarrow +\infty$. The advantage of introducing this mild non-locality is that the ground state of $H$ has a UV-\emph{finite} bipartite entanglement entropy in this basis (while the standard entanglement entropy is UV-divergent) \cite{Tilloy:2022kcn}.

We now define the two-field RCMPS ansatz as: 
\begin{equation}\label{eq:RCMPS_two_species} 
  |Q, R_1, R_2\rangle = \text{tr}\left[\mathcal{P}\e^{\int_{\mathbb{R}} \upd x Q + R_1\hat{a}_1^\dagger(x) + R_2\hat{a}_2^\dagger(x)} \right]|0\rangle,
\end{equation}
where $Q, R_1, R_2$ are $D\times D$ complex matrices containing the variational parameters, $D$ is the \emph{bond dimension} (controlling the expressiveness of the ansatz), $\mathcal{P} \exp$ is the path-ordered exponential, the trace is taken over the matrix space, and $|0\rangle$ is the Fock vacuum, \textit{i.e.} $\forall x, \; a_j(x)|0\rangle=0$. 

Shifting $Q$ by a constant merely changes the norm of the state, and we may thus fix the latter to $1$ in the thermodynamic limit. Further, the state is invariant under the `gauge' transformation: $Q \rightarrow X^{-1}QX,\, R_j \rightarrow X^{-1}R_jX$, where $X \in GL(D,\mathbb{C})$. This can be exploited to reduce the number of variational parameters by casting the state into a \emph{left canonical form} \cite{PhysRevLett.104.190405, Haegeman2013} defined by $Q + Q^{\dagger} + \sum_j R^{\dagger}_jR_j = 0$. This allows us to write $Q = -iK - \sum_j R^{\dagger}_jR_j/2$ where $K$ is Hermitian, a more parsimonious parameterization we now use. We call $\mrcmps$ the space of $\ket{K,R_1,R_2}$ for fixed bond dimension $D$.

\begin{figure}[t]
    \includegraphics[width=1.2\linewidth]{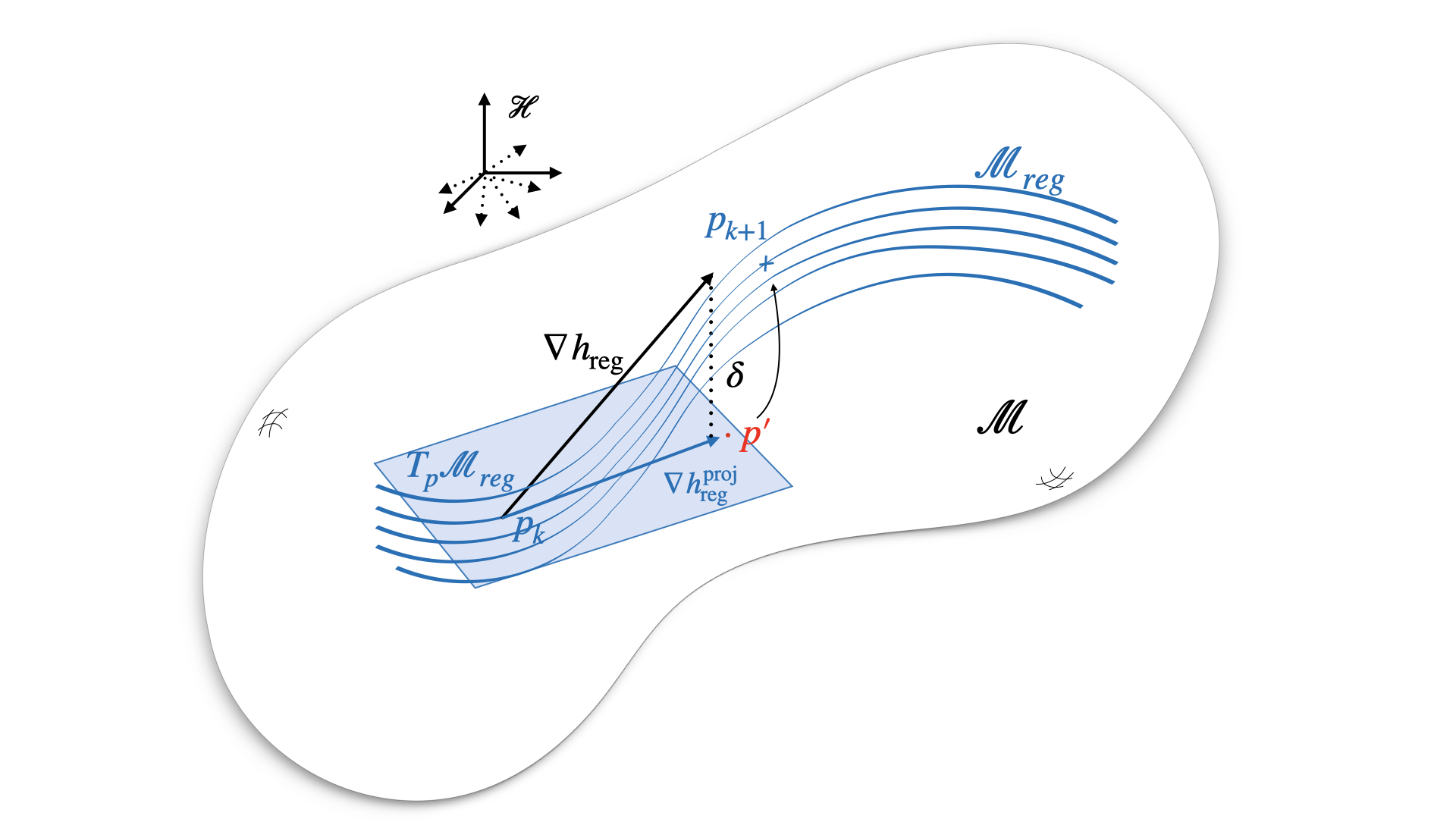}
    \caption{Representation of the RCMPS manifold $\mrcmps$, parameterized by $R_1$, $R_2$, and $Q$ matrices, as a subspace in the full Hilbert space $\mathscr{H}$ of the field theory. The restricted manifold of commuting $R_1$ and $R_2$ matrices is denoted by $\mathcal{M}_{\text{reg}}$.}
    \label{fig:two_species}
\end{figure}

The expectation value of the Hamiltonian density $\hat{h}$ on a state $p=\ket{K,R_1,R_2}$ can in principle be computed using the same techniques as in the single field case \cite{Tilloy:2021yre, Tilloy:2021hhb}, apart from two minor technical subtleties we discuss in the End Matter \ref{app:multi_field}. However, as we argued before, the space $\mrcmps$ happens to be too large, and contains states that are too irregular for the Hamiltonian we consider. Indeed, the expectation value of the free Hamiltonian density $\langle \hat{h}_{1}^{0}+\hat{h}_{2}^{0}\rangle$ contains a divergent contribution $h_\text{div}$~(see End Matter \ref{app:regularity}):
\begin{equation}
\begin{split}
    &h(p):= \langle K,R_1,R_2 | \hat{h}|K,R_1,R_2\rangle  = h_\text{reg}(p) + h_\text{div}(p)~,\\
    &\text{with} \quad h_\text{div} = 4\tr\left( [R_1,R_2]^\dagger [R_1,R_2] \rho_0\right )\underset{+\infty}{\underbrace{\int_\mathbb{R} \upd x \,  J^2(x)}}~, \nonumber
\end{split}
\end{equation} 
where $h_\text{reg}(p)$ is finite, $\rho_0$ is the positive trace $1$ matrix such that $Q\rho_0 + \rho_0 Q^\dagger + \sum_k R_k\rho_0 R^\dagger_k=0$, and the kernel $J(x)$ diverges as $|x|^{-1/2}$ near $x=0$. Hence, to ensure a finite energy density, we must enforce $[R_1, R_2] = 0$, in which case $h_\text{div}(p)=0$, and the energy density is well defined. We thus restrict the ground state search to $\mathcal{M}_{\text{reg}} = \{\ket{K, R_1, R_2} \mid [R_1, R_2]=0\}$ (see Fig. \ref{fig:two_species}).

\emph{Constrained Optimization} --- 
To find a variational approximation to the ground state we need to find: 
\begin{equation}\label{eq:Optimisation_prob}
  \argmin_{p \in \mrcmps} h(p) = \argmin_{p\in \mreg} h_\text{reg}(p)  \, .
\end{equation}
Such a constrained optimization problem could in principle be solved in many ways, \textit{e.g.} by introducing a Lagrangian, or by adding a penalty for constraint violation. We would like to keep a strictly variational approach, providing rigorous energy density upper bounds. This requires having a point $p_k$ satisfying the constraint exactly at each iterate of our algorithm. To this end,  we view $\mreg$ as a submanifold of $\mrcmps$, which is itself a submanifold of the full Hilbert space $\mathscr{H}$, and iteratively lower the energy with a Riemannian gradient descent algorithm. The advantage of this approach is that one can reuse the tools (tangent space, metric, gradient) developed for the single field case, adding only two new geometric tools (tangent projection and retraction).

Let us first define the required geometric tools on the non-regular manifold $\mrcmps$. The tangent vector at a point $p$ is \emph{a priori} parametrized by complex matrices $V,W_1,W_2$:
\begin{equation}
\begin{split}
    |V,W_1,W_2\rangle_{p} = \sum_{\alpha,\beta = 1}^D&\Bigg(V^{\alpha\beta}\frac{\delta }{\delta Q^{\alpha\beta}} \\
    +& \sum_{j=1}^2 W_j^{\alpha\beta}\frac{\delta }{\delta R_j^{\alpha\beta}}\Bigg) \ket{Q,R_1,R_2}.
\end{split}
\end{equation}
However, gauge freedom again implies that some of these parameters are redundant, \textit{i.e.} lead to the same tangent vector.  Working in the left canonical gauge, one can fix $V = -\sum_j R^{\dagger}_{j}W_{j}$ \cite{Haegeman2013,vanderstraeten2019_tangentspace}. The tangent space $T_p\mrcmps$ is thus parameterized by the pair of matrices $W=(W_1,W_2)$ and we write the corresponding tangent vectors $|W\rangle_p$.

The manifold $\mrcmps$ has a natural Riemannian metric $\mathsf{g}$ on its tangent space: the Hilbert space inner product~\cite{vanderstraeten2019_tangentspace, Haegeman2013}. One can show that the induced metric has a simple expression directly in terms of $W$:
\begin{equation}\label{eq:metric}
\begin{split}
  \mathsf{g}_p(\Tilde{W},W) &:= \text{Re}\langle\Tilde{W}\ket{W}_{p}\\
                            &=\text{Re} \left\{\tr\left[\left(\Tilde{W}^\dagger_{1} W_1 + \Tilde{W}^\dagger_{2} W_2 \right)\rho_{0}(p)\right]\right\}~.
\end{split}
\end{equation}
The gradient is defined from the infinitesimal change of the (regular part of the) energy density in the direction of a tangent vector:
\begin{equation}
  h_\text{reg}(p + \varepsilon |W\rangle_p) = h_\text{reg}(p) + \varepsilon \, \mathsf{g}_p(\nabla h_\text{reg} , W) + o(\varepsilon) \, .
\end{equation}
Given the metric $\mathsf{g}$ defined in \eqref{eq:metric}, and an expression for $h_\text{reg}(p)$, one can compute $\nabla h_\text{reg}$ with the same asymptotic cost as $h_\text{reg}(p)$ (\textit{i.e.} $\mathcal{O} (D^3)$) using the \emph{backward differentiation} method used for the single field case in~\cite{Tilloy:2021yre, Tilloy:2022kcn}. 

We now need to define a gradient descent algorithm that stays on the manifold $\mreg$. This requires two new geometric ingredients. We start from an initial $p_0 \in \mreg$, and at each new iterate $p_k$ we need i) the orthogonal projection of the gradient onto $T_{p_k}\mreg$ and ii) \emph{a} retraction map $\text{Ret}_{p_k}:T_{p_k}\mreg \rightarrow \mreg$ giving us a new point $p_{k+1}$.

\emph{1. Gradient Projection:}
To project the gradient, we first need to characterize the tangent space $\mreg$. It is obtained by taking the first order variation of the constraint
\begin{equation}\label{eq:tangent_constraint}
T_p\mreg
=\Bigl\{(W_1,W_2):\ [W_1,R_2]+[R_1,W_2]=0\Bigr\}.
\end{equation}
In other words, $T_p\mreg$ is the kernel of the linear map $L(W_1,W_2) = [W_1,R_2]+[R_1,W_2]$. 

To define the orthogonal projection onto $T_p\mreg$, we parameterize the orthogonal complement of the kernel of $L$, which is given by the range of $L^\dagger$, \textit{i.e.} $T_p\mreg^\perp = (\text{ker}L)^\perp = \text{im}L^\dagger$. For all $\Lambda \in \mathbb{C}^{D\times D}$, we have $\tr\left[L(W)^\dagger\Lambda\right]= \mathsf{g}_p\left[W,L^\dagger(\Lambda)\right]$, and the range of $L^\dagger$ is thus parametrized by pairs $\delta = (\delta_1,\delta_2)$ of the form
\begin{equation}
\delta = \bigl(-\,[R_2^\dagger,\Lambda]\rho_0^{-1},\ [R_1^\dagger,\Lambda]\rho_0^{-1}\bigr)~,
\end{equation}
where $\Lambda$ is any $D\times D$ complex matrix.
We may now decompose $\nabla h_\text{reg} = \nabla h_\text{reg}^\text{proj} + \delta$ with $\nabla h_\text{reg}^\text{proj}\in T_p\mreg$ and $\delta \in T_p\mreg^\perp$. 

The vector $\nabla h_\text{reg}^\text{proj}$ belongs to the regular tangent space if the constraint \eqref{eq:tangent_constraint} is verified:
\begin{equation}
  [R_1, (\nabla h_\text{reg}^\text{proj})_2 ] + [(\nabla h_\text{reg}^\text{proj})_1,R_2] = 0\, .
\end{equation}
We can now inject $\nabla h_\text{reg}^\text{proj} =  \nabla h_\text{reg} - \delta(\Lambda)$ to obtain 
\begin{equation} \begin{split}
  [R_1,[R_1^\dagger,\Lambda]\rho_0^{-1}]+&[R_2,[R_2^\dagger,\Lambda]\rho_0^{-1}]\\
  = [&R_1,(\nabla h_\text{reg})_2 ]+[R_2,(\nabla h_\text{reg})_1]
~,
  \label{eq:PG_rho}
\end{split}
\end{equation}
which is a linear equation in $\Lambda$, of the form $\mathcal{A}(\Lambda) = b$.

Equation~\eqref{eq:PG_rho} can be solved efficiently using iterative methods, \emph{e.g.} the generalized minimal residual method (GMRES). Solving for $\Lambda$ determines $\delta$ and hence
$\nabla h_\text{reg}^\text{proj}$. By construction, it is the orthogonal projection of the gradient onto $T_p\mreg$ with respect to the metric $\mathsf{g}$, and hence the steepest descent direction at $p\in\mreg$.  

\emph{2. Retraction:}
Once we have the projected gradient $\nabla h_\text{reg}^\text{proj} \in T_p\mreg$, it is tempting to do a trial step of size $\alpha$  in the direction opposite to the gradient $ R_1' = R_1 - \alpha (\nabla h_\text{reg}^\text{proj})_1 , R_2' = R_2 - \alpha (\nabla h_\text{reg}^\text{proj})_2 $, and
\begin{equation}
  K \longmapsto K'(\alpha) = K - \frac{i\alpha}{2}\sum_{j=1}^2(W_j^\dagger R_j - R^\dagger_j W_j)~,
\end{equation}
which preserves the gauge. 
This trial step stays on $\mreg$ to leading order in $\alpha$. However, it has no reason to remain on $\mreg$ once $\alpha$ is not infinitesimal. Indeed $[R_1',R_2']=\alpha^2[(\nabla h_\text{reg}^\text{proj})_1,(\nabla h_\text{reg}^\text{proj})_2 ] \neq 0$ in general.

For our Riemannian gradient descent algorithm, we need a retraction map that stays on $\mreg$ for \emph{all} $\alpha$ while agreeing with this naive trial step to first order in $\alpha$. Intuitively, since $R_1'$ and $R_2'$ commute to first order in $\alpha$, we will not change the leading order by forcing them to commute. One option is to take $R_1'$ as reference, and diagonalize it $R'_1 = V D'_1 V^{-1}$. We may then express $R_2'$ in the eigenbasis of $R_1'$ and simply discard the off-diagonal components
\begin{equation}\nonumber
\begin{split}
\widetilde R_2
    = V\,\mathrm{diag}(V^{-1} R_2' V)\,V^{-1}~.
\end{split}
\end{equation}
Since $R_1'$ and $\widetilde R_2$ are diagonal in the same
basis, they commute exactly. The correction to $R_2'$ is of order $\mathcal O(\alpha^2)$ and does not affect the $\mathcal{O}(\alpha)$ change, which remains opposite to the gradient. The state $\ket{K',R'_1,\widetilde{R}_2}$ now belongs to $\mreg$ for any value of $\alpha$. In practice, one need not take $\alpha$ small and one can find a reasonable value at each iteration using a backward line search.

The two steps described above rely on linear algebraic operations scaling as $\mathcal{O}(D^3)$ (assuming a small Krylov subspace for GMRES). Computing the energy density gradient also costs $\mathcal{O}(D^3)$ in typical CMPS and RCMPS problems. However, for RCMPS, this asymptotic cost hides a huge pre-factor. Indeed, the $\mathcal{O}(D^3)$ is associated to each evaluation of the generator of an ODE, which needs to be evaluated several thousand times to get a precise solution~(see End Matter \ref{app:multi_field}). Consequently, the computational bottleneck of the algorithm lies in the evaluation of the energy and its gradient, rather than on the geometric steps we described, which need not be implemented optimally. This would be different if the same method were applied to CMPS, where the energy density is much cheaper to evaluate.

\begin{figure}[t]
    \includegraphics[width=\columnwidth]{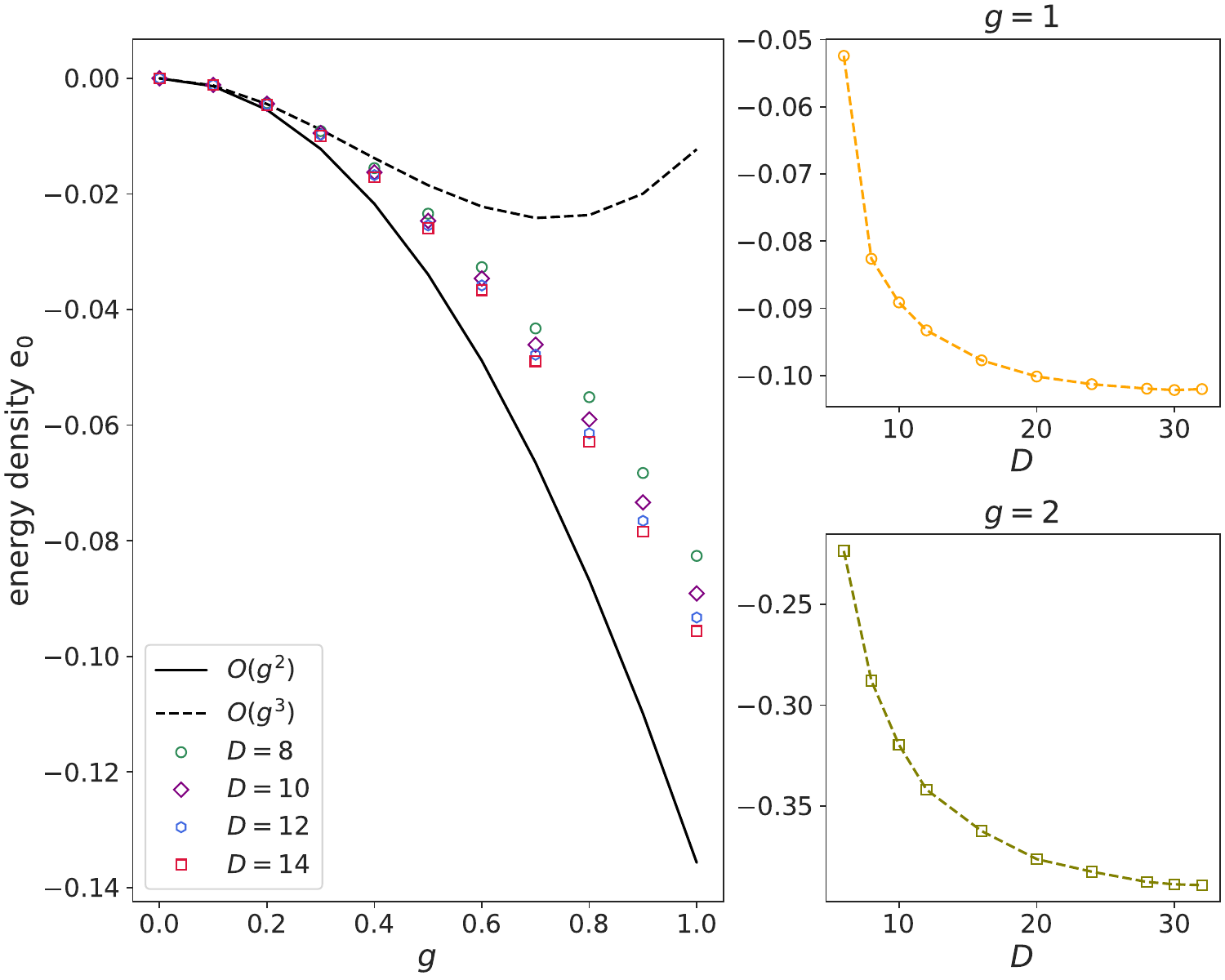}
    \caption{Ground state energy density $e_0$ for $g=\lambda$. RCMPS results (markers) show excellent agreement with perturbation theory at order $g^2$ and $g^3$ at weak coupling. Already for moderate coupling $g\geq 0.3$, the RCMPS results are far more precise, even at the lowest $D$ we consider.}
    \label{fig:pert_comparison}
\end{figure}

\emph{Results}---
We evaluate the expectation values and their gradients using \texttt{DifferentialEquations.jl}~\cite{Rackauckas-2017}, solve the linear problem \eqref{eq:PG_rho} associated to tangent space projection using \texttt{KrylovKit.jl}~\cite{KrylovKitRepo}, and finally, carry the gradient descent with \texttt{OptimKit.jl}~\cite{OptimKit}, a package designed to perform optimization on generic manifolds. After a few hundred to a few thousand iterations (depending on the physical parameters $g,\lambda$ and bond dimension $D$), we obtain an optimal RCMPS approximation to the ground state of $H$.

Since the model we consider is not exactly solvable, we first validate our method in the weak-coupling regime by comparing our estimate of the ground state energy density $e_0$ with perturbation theory. Figure \ref{fig:pert_comparison} shows the results for the $g=\lambda$ line. We observe that the RCMPS results match perturbation theory as $g \to 0$, a crucial sanity check. However, even the lowest bond dimensions we consider are considerably more accurate than perturbative results, already for moderate values of $g \geq 0.3$.

\begin{figure}[t]
  \centering
  \includegraphics[width=\columnwidth]{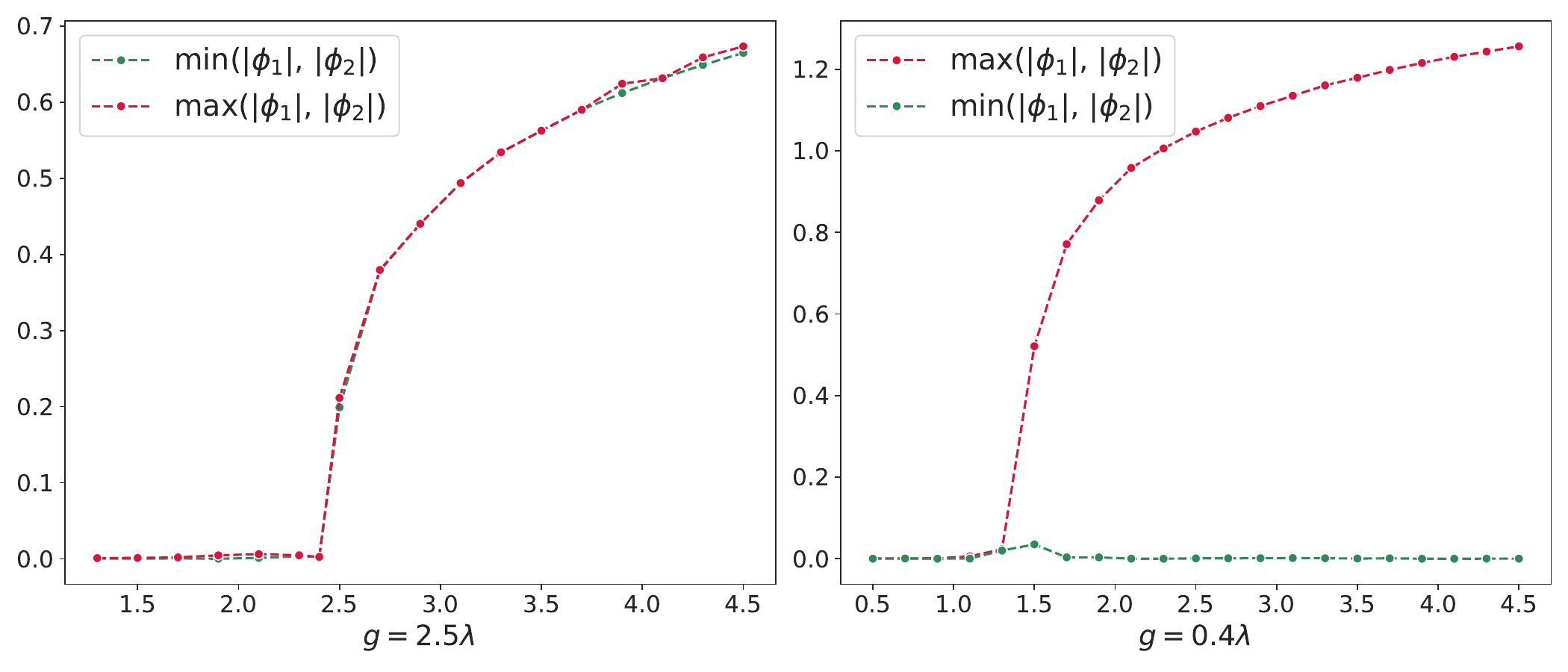}
  \caption{Order parameters $\langle\hphi_1\rangle$ and $\langle\hphi_2\rangle$ across the phase transition for
  (a) $g > \lambda$ and
  (b) $\lambda > g$, for $D=18$.
  The method correctly identifies the distinct symmetry-breaking patterns.}
  \label{fig:phase_diagram}
\end{figure}

The model gets more interesting in the non-perturbative regime. The potential \eqref{eq:potential} is invariant under $\hphi_k \leftrightarrow -\hphi_k,\; \hphi_1 \leftrightarrow \hphi_2$. This corresponds to the dihedral group $D_4$, \textit{i.e.} the discrete symmetry group of a square. It can be broken spontaneously to any of its subgroups~\footnote{The $D_4$ group has 10 subgroups, out of which 8, excluding the identity element and the full group itself, are non-trivial.}, giving different phases where the corresponding order parameters acquire a vacuum expectation value (vev). Our method is able to accurately capture some of them. We consider the simplest order parameters $\langle\hphi_1\rangle$ and $\langle\hphi_2\rangle$ for two different cuts in the $g-\lambda$ parameter space shown in Fig.~\ref{fig:phase_diagram}. In the regime $g > \lambda$, we observe a phase transition in which both fields acquire an equal vev, signaling that the $D_4$ symmetry is broken into a residual $\mathbb{Z}_2$ characterised by $\hphi_1 \leftrightarrow  \hphi_2$. For $\lambda > g$, we observe a transition to a different phase where one field acquires a vev while the other remains zero. This captures $D_4$ breaking to a \emph{different} $\mathbb{Z}_2$ described by $\hphi_k\leftrightarrow-\hphi_k$ for \emph{only} one of the $k$s. The method clearly distinguishes these different symmetry-breaking patterns, which one could not obtain by \textit{e.g.} taking a product of single-field RCMPS.

\begin{figure}
    \centering
    \includegraphics[width=\columnwidth]{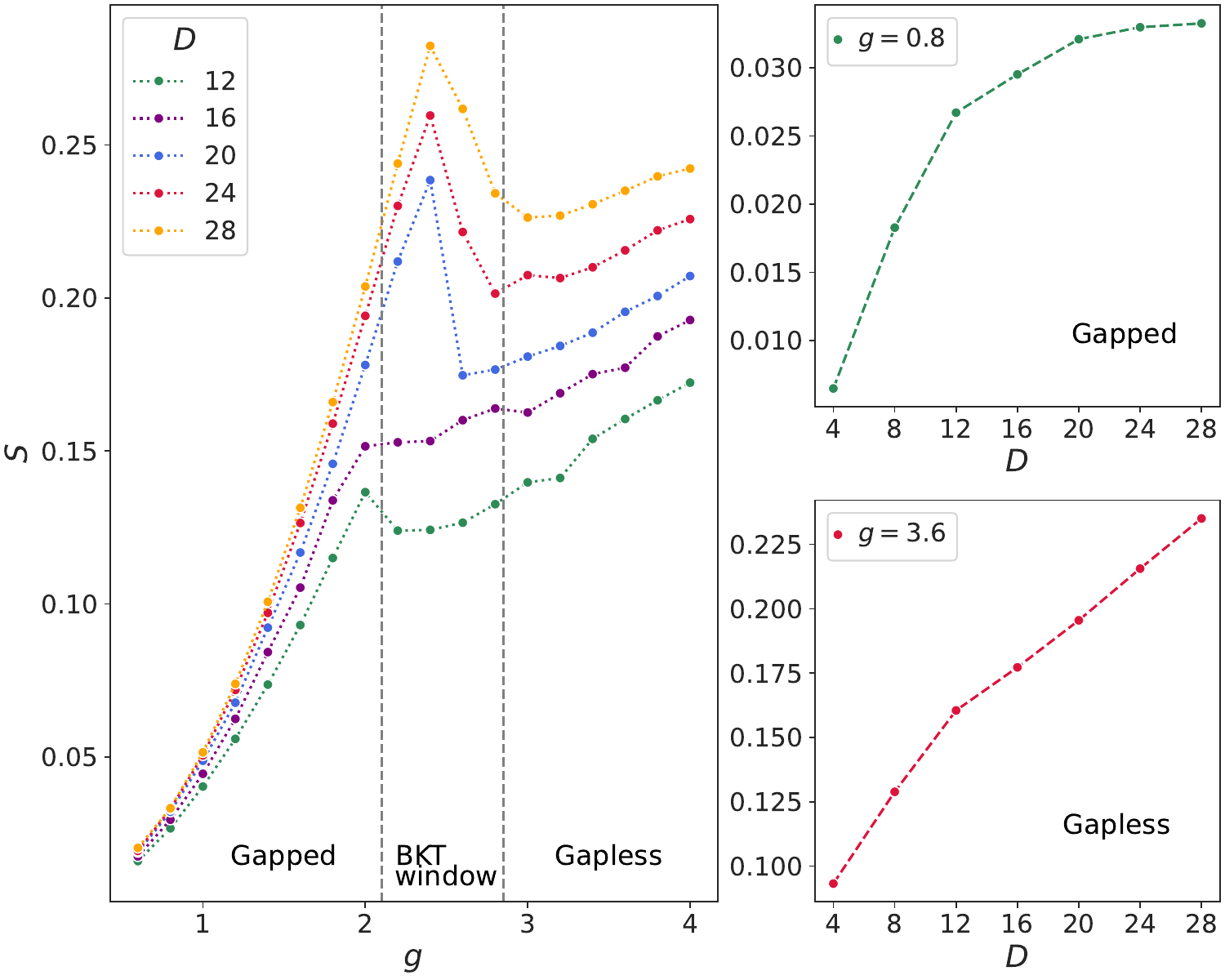}
    \caption{RCMPS entanglement entropy $S$ for $g= \lambda$ as a function of $g$ for different bond dimensions $D$.}
    \label{fig:entropy_o2}
\end{figure}
Finally, when $\lambda=g$, the dihedral symmetry is enhanced to the continuous $O(2)$ group. On this line, spontaneous breaking of the continuous symmetry is forbidden \cite{PhysRevLett.17.1133, PhysRev.158.383, Coleman:1973ci}, and the system is expected to undergo a BKT \cite{Kosterlitz:1973xp} transition into a critical phase. The RCMPS bipartite entanglement entropy \footnote{This is a distinct notion of entanglement entropy than the usual bipartite case where the state is `cut' locally. Here, we are partitioning the full Hilbert space into products of factors corresponding to $\hat{a}(x)$, which disentangles the free Fock vacuum fully. The RCMPS entanglement entropy captures the \emph{extra} entropy introduced by the interactions in the Hamiltonian. For a discussion, see \cite{Tilloy:2022kcn}.} defined as $S = -\text{tr}(\rho_0 \log \rho_0)$ serves as a probe for this transition. In a gapped phase, this entropy is expected to be finite and should converge as $D$ increases, while in the gapless case, it should grow algebraically as $D$ is increased.

We indeed observe a peak in $S$ near the expected transition point, beyond which the entropy fails to converge with increasing bond dimension $D$ (see Fig~\ref{fig:entropy_o2}). 
This persistent gap in the right part of Fig.~\ref{fig:entropy_o2} is a characteristic signature of the gapless phase. To locate the transition point more accurately, and finely describe the critical phase, one would need to carefully extrapolate the results in $D$ with \emph{finite entanglement scaling} \cite{tagliacozzo2008entanglement,pollmann2009_finiteentanglementscaling,stojevic2015entanglementscaling}.

\emph{Conclusion}---
We presented a variational framework for multi-field QFTs using RCMPS. We turned the ground-state search into a well-posed Riemannian optimization problem and dealt with the long standing problem of regularity constraint $[R_1, R_2]=0$ with a combination of gradient projection and retraction. We applied the method to a $D_4$-symmetric scalar field example. This allowed us to accurately capture non-perturbative parts of the phase diagram, with subtle symmetry breaking patterns and a topological phase transition on the $O(2)$ symmetric line. To our knowledge, no similar results for a two-field model have yet been obtained with alternative non-perturbative methods (\textit{e.g.} lattice MPS \cite{milsted2013_MPSforPhi4}, Monte Carlo \cite{bosetti2015montecarlo,bronzin2019montecarlo}, or Hamiltonian truncation \cite{Hogervorst:2014rta,rychkov2015,elias-miro2017NLOrenormalization}, which have all already been used for a single field). 

An immediate future direction is to fully characterize the phase diagram of the model, and more accurately pin the location of the BKT transition. This will require an extension of finite entanglement scaling techniques to RCMPS, and ideally a way to include global symmetries into the ansatz, as was done with MPS. The optimization algorithm we have proposed can be straightforwardly generalized to $n$ species and hence can already be used to study scalar field systems invariant under the \emph{hyperoctahedral group} $B_n = (\mathbb{Z}_2)^n\rtimes S_n$ as well as $O(n)$ groups, which are not directly solvable. It is also a crucial first step in the RCMPS study of problems with Dirac fermions, which have a similar regularity constraint. The ability to work with multi-field RCMPS also finally makes it possible to compute the expectation values of general non-Gaussian continuous tensor network state in $2$ space dimensions~\cite{tilloy2019ctns}. This opens the way to the use of continuous tensor networks beyond $1+1$ dimensions.

\begin{acknowledgments}
\emph{Acknowledgements.}---
We thank Jutho Haegeman, Wei Tang, and Beno\^{i}t Tuybens for valuable discussions. We became aware of their independent work on multiple scalar fields using CMPS while this project was in progress, and we benefited from subsequent exchanges. We are also grateful to Edoardo Lauria and Sophie Mutzel for helpful discussions on various aspects of this project. We are supported by the European Union (ERC, QFT.zip project, Grant Agreement no. 101040260). The computations were carried out using the CLEPS cluster managed by Inria, Paris.
\end{acknowledgments}
\bibliography{main}

\appendix

\section*{End Matter}

\refstepcounter{appsec}%
\section{\theappsec. Derivation of Expectation Values via Generating Functionals\label{app:multi_field}}

We start by recalling how local field expectation values of the form $\langle:\hat{\phi}^n(x):\rangle$, that appear in the potential, are computed for the single field case (see \cite{Tilloy:2021hhb, Tilloy:2021yre} for more details).

\subsection{Single-Field Formalism}

For a single scalar field, the generating functional $\mathcal{Z}_{j',j}$ for sources $j'(x)$ and $j(x)$ is defined as:
\begin{equation}
\begin{split}
    \mathcal{Z}_{j',j} = \langle Q,R | \exp&\left[\int \upd x \, j'(x)\hat{a}^\dagger(x)\right] \\
    &\times\exp\left[\int \upd x \, j(y)\hat{a}(y)\right] | Q,R \rangle~
\end{split}
\end{equation}
Using the Baker-Campbell-Hausdorff (BCH) formula, and the fact that the RCMPS creation/annihilation operators $\hat{a}^\dagger, \hat{a}$ satisfy canonical commutation relations $[a(x),a^{\dagger}(y)] = \delta (x-y)$, we can rewrite the functional as:
\begin{equation}
    \begin{split}
    \mathcal{Z}_{j',j} &= \langle Q, R+j I | Q, R+j^* I\rangle \exp\left(-\int \upd x \, j'(x)j(x)\right),\\
    &= \text{tr}\left[\mathcal{P}\exp\left(\int \upd x \left( \mathbb{T} + j'(x)R \otimes I + j(x)I \otimes \overline{R}\right)\right)\right]~.
    \end{split}
\end{equation}
where $\mathbb{T} = Q \otimes I + I\otimes \overline{Q} + R\otimes \overline{R}$ is the transfer matrix associated with the state $|Q,R\rangle$.

To compute local observables like $\langle:\!\hat{\phi}^n(x)\!:\rangle$, we consider the expectation value of the vertex operator $\langle V_\beta(x) \rangle = \langle :\!e^{\beta\hat{\phi}(x)}\!: \rangle$. Using the relation $\hat{\phi}(x) = \int \upd y \, J(x-y)[\hat{a}(y)+\hat{a}^\dagger(y)]$, the vertex operator expectation value becomes a specific instance of the generating functional with sources $j(y)=j'(y)=\beta J(x-y)$, which can be expressed as the solution of an ordinary differential equation (ODE). 

We write the ODE in the superoperator representation, defined via the isomorphism $\ket{w}\otimes \ket{v} \cong \ket{w}\bra{v}$, to exploit the fact that the superoperator $\mathcal{L}$ corresponding to $\mathbb{T}$ acts via left/right matrix multiplication on  $\mathbb{C}^{D\times D}$ matrices in contrast to $\mathbb{T}$ which is a linear map on a $\mathbb{C}^{D^2}$ vector space. Explicitly, $\mathbb{T}\ket{w}\otimes \ket{v} \longmapsto\mathcal{L}(\ket{w}\bra{v}) = Q\ket{w}\bra{v} + \ket{w}\bra{v}Q^\dagger + R\ket{w}\bra{v}R^\dagger$, which can be evaluated at cost $D^3$, compared to $D^4$ for the naive action with $\mathbb{T}$. With this rewriting we have:
\begin{equation}
    \langle V_\beta(x) \rangle = \lim_{y\to\infty} \text{tr}[\rho_\beta(y)],
\end{equation}
where $\rho_\beta(y)$ is the solution of the ODE 
\begin{equation}
\begin{split}
    &\frac{d\rho_\beta(y)}{\upd x} = \mathcal{L}\cdot\rho_\beta(y) + b J(x-y) \left(R\rho_\beta(y) + \rho_\beta(y)R^\dagger\right),\\
    &\lim_{y\to-\infty}\rho_\beta(y)=\rho_0.
\end{split}
\end{equation}
The desired field monomials are then extracted by differentiation: $\langle :\hat{\phi}^n(x): \rangle = \left. \frac{\partial^n}{\partial \beta^n} \langle V_\beta(x) \rangle \right|_{\beta=0}$. Doing the forward differentiation explicitly yields a hierarchy of linear ODEs for the derivatives $\rho^{(k)}(y) := \left. \frac{\partial^k \rho_\beta(y)}{\partial \beta^k} \right|_{\beta=0}$:
\begin{equation}
\begin{split}
    \frac{d\rho^{(k)}(y)}{\upd x} = \mathcal{L}\cdot\rho^{(k)}(y) + k J(x-y) \Big(R\rho^{(k-1)}(y) \\
        + \rho^{(k-1)}(y)R^\dagger\Big).
\end{split}
\end{equation}
This non-autonomous system of linear ODEs can be solved efficiently using simple ODE solvers, for example explicit Runge-Kutta schemes, implemented in the \texttt{DifferentialEquations.jl} Julia package~\cite{Rackauckas-2017}. The asymptotic cost is that of applying the generator, \textit{i.e.} $\mathcal{O}(D^3)$.

\subsection{Multi-Field Formalism}
To compute expectation for a RCMPS with multiple fields, we follow the same philosophy.The derivation of expectation values involving a single field is identical up to the replacement of $\mathcal{L}$ (equivalently $\mathbb{T}$) by its natural two-field generalization
\begin{equation}
  \mathcal{L}(\rho) := Q\rho + \rho Q^\dagger + R_1 \rho R_1^\dagger + R_2 \rho R_2^\dagger\,.
\end{equation}
To compute the expectation values of mixed field operators, such as $\langle:\hat{\phi}_1^m \hat{\phi}_2^n:\rangle$, we can introduce two-parameter vertex operator
\begin{equation}
    \langle V_{\alpha\beta} \rangle = \langle :e^{\alpha\hat{\phi}_1(x)}::e^{\beta\hat{\phi}_2(x)}: \rangle.
\end{equation}
Using the fact that the two fields $\hat{\phi}_1$ and $\hat{\phi}_2$ commute, the expectation value $\langle V_{\alpha\beta} \rangle$ can be computed in a manner similar to the single field case. We have $\langle V_{\alpha\beta} \rangle = \lim_{x\to\infty} \text{tr}[\rho_{\alpha\beta}(x)]$ where $\rho_{\alpha\beta}(x)$ solves the system of ODEs
\begin{equation}
\begin{split}
    \frac{\upd\rho_{\alpha\beta}(x)}{\upd x} = \mathcal{L}\cdot\rho_{\alpha\beta}(x) + J(x)\Big[ \alpha(R_1\rho_{\alpha\beta}(x) \\
    + \rho_{\alpha\beta}(x)R_1^\dagger) + \beta(R_2\rho_{\alpha\beta}(x) + \rho_{\alpha\beta}(x)R_2^\dagger)\Big],    
\end{split}
\end{equation}
with the initial condition $\lim_{x\to-\infty} \rho_{\alpha\beta}(x) = \rho_0$, where $\rho_0$ is the trace 1 matrix such that $\mathcal{L}(\rho_0)$.

The expectation value of a normal-ordered monomial is obtained by differentiation:
\begin{equation}
    \langle :\hat{\phi}_1^m \hat{\phi}_2^n: \rangle = \left. \frac{\partial^{m+n}}{\partial\alpha^m \partial\beta^n} \langle V_{\alpha\beta} \rangle \right|_{\alpha,\beta=0}.
\end{equation}
Forward differentiating explicitly leads to a hierarchy of linear ODEs for the derivatives $\rho^{(m,n)}(x) := \left. \frac{\partial^{m+n}\rho_{\alpha\beta}(x)}{\partial\alpha^m \partial\beta^n} \right|_{\alpha,\beta=0}$
  \begin{equation}
\begin{split}
    &\frac{\upd \rho^{(m,n)}}{\upd x} = \mathcal{L}\cdot\rho^{(m,n)} + m J(x) \Big(R_1\rho^{(m-1,n)}\\
    &+ \rho^{(m-1,n)}R_1^\dagger\Big) + n J(x) \left(R_2\rho^{(m,n-1)} + \rho^{(m,n-1)}R_2^\dagger\right).
\end{split}
\end{equation}
with initial condition $\rho^{(0,0)}(x) = \rho_0$, all the other matrices being zero. The desired expectation value is then $\langle :\hat{\phi}_1^m \hat{\phi}_2^n: \rangle = \lim_{x\to\infty} \text{tr}[\rho^{(m,n)}(x)]$.

\refstepcounter{appsec}%

\section{\theappsec. Kinetic Energy and Regularity Condition\label{app:regularity}}
The kinetic energy density for the field $j$ can be written in terms of the position modes $a_j(x)$:
\begin{equation}
\begin{split}
    \langle \hat{h}_j^0 \rangle = 2m^2\left\langle\left(\int \upd x\, J(x)\hat{a}_j^\dagger(x)\right)\left(\int \upd y\, J(y)\hat{a}_j(y)\right)\right\rangle\\
    + 2\left\langle\left(\int \upd x \,J(x)\partial_x \hat{a}_j^\dagger(x)\right)\left(\int \upd y \, J(y)\partial_y \hat{a}_j(y)\right)\right\rangle.
\end{split}
\end{equation}
The term involving derivatives requires careful treatment. The action of $\partial_x \hat{a}_j(x)$ on an RCMPS state is found by differentiating the action of $\hat{a}_j(x)$:
\begin{equation}
\begin{split}
    &\partial_x \hat{a}_j(x) |Q,R\rangle = \text{tr}\Bigg[\hat{U}(-\infty,x) \Bigg( [Q, R_j]\\
    & \qquad \qquad \qquad \qquad + \sum_k [R_k, R_j] \hat{a}_k^\dagger(x) \Bigg) \hat{U}(x,\infty) \Bigg] |0\rangle~,\\
    &\text{with} \;\, \hat{U}(a,b):= \mathcal{P}\exp\left[\int_a^b \upd y\left(Q + \sum_k R_k \hat{a}^\dagger_k (y) \right) \right]~.
\end{split}
\end{equation}
When computing the expectation value $\langle \partial_y \hat{a}_j^\dagger(y) \partial_x \hat{a}_j(x) \rangle$, the term $[R_k, R_j]\hat{a}_k^\dagger(x)$ gives a Dirac delta contribution:
\begin{equation}
\begin{split}
&\big\langle \partial_y \hat a_j^\dagger(y)\,\partial_x \hat a_j(x)\big\rangle 
= (\text{smooth terms})(x-y) \\ 
&+\delta(x-y)\,
\sum_{k} \tr \left([R_j,R_k]^\dagger [R_j,R_k] \rho_0\right).
\end{split}
\end{equation}
The kinetic energy density $\langle \hat{h}_j^0 \rangle$ thus contains a divergent contribution. Summing the contribution from the kinetic energy density of all the fields, we get the divergent part
\begin{equation}
\begin{split}
     &h_\text{div} = 2 \sum_{k,j} \tr \left([R_j,R_k]^\dagger [R_j,R_k] \rho_0\right)\int \upd x \, J(x)^2
\end{split}
\end{equation}
For $x$ small, $J(x) \sim |x|^{-1/2}$ so $J(x)^2 \sim |x|^{-1}$. The integral $\int \upd x J(x)^2$ is therefore logarithmically divergent at the origin. To obtain a finite and physically meaningful energy density, this divergent term must vanish for all pairs of fields. This imposes the regularity condition:
\begin{equation}
    [R_j, R_k] = 0 \quad \forall j, k.
\end{equation}
This constraint is essential for the variational problem to be well-posed for any multi-component RCMPS.

\end{document}